# Large Magnetoresistance at Room Temperature in Ferromagnet/Topological Insulator Contacts

S. Majumder[*], S. Guchhait, R. Dey, L. F. Register, and S. K. Banerjee,

**Abstract:**
   We report magnetoresistance for current flow through iron/topological insulator (Fe/TI) and Fe/evaporated-oxide/TI contacts when a magnetic field is used to initially orient the magnetic alignment of the incorporated ferromagnetic Fe bar, at temperatures ranging from 100 K to room temperature. This magnetoresistance is associated with the relative orientation of the Fe bar magnetization and spin-polarization of electrons moving on the surface of the TI with helical spin-momentum locking. The magnitude of the observed magnetoresistance is relatively large compared to that observed in prior work.

* Corresponding author: e-mail sarmita@utexas.edu.

## I. Introduction

Topological insulators (TI) are insulating in the bulk and have a Dirac-like two-dimensional (2D) surface- state band structure. An essential feature of these TI surface states is helical spin-momentum locking due to strong spin-orbit interactions, making these a promising material for spintronic applications [1-3]. An applied charge current induces a net spin polarization/accumulation at the TI surface, which can be controlled by the amplitude and the polarity of the current [4-15]. For example, electron flow in the negative $x$ direction within the surface state conduction or valence band and with the $z$ axis pointing away from the surface results in a net electron spin in the positive $y$ direction.

Manuscript is submitted on February 19, 2016 in IEEE Trans. This work was supported in part by the NRI SWAN program, the NSF Nanosystems Engineering Research Center (NERC) for Nanomanufacturing Systems for Mobile Computing and Mobile Energy Technologies (NASCENT), and the NSF NNCI program.

   S. Majumder is with the Microelectronics Research Center (MRC), University of Texas, Austin, TX 78758 USA, (sarmita@utexas.edu)
   R. Dey is with the Microelectronics Research Center (MRC), University of Texas, Austin, TX 78758 USA, (rikdey@utexas.edu)
   L. F. Register is with the Microelectronics Research Center (MRC), University of Texas, Austin, TX 78758 USA, (register@austin.utexas.edu)
   S. K. Banerjee is with the Microelectronics Research Center (MRC), University of Texas, Austin, TX 78758 USA, (banerjee@ece.utexas.edu)
   S. Guchhait was with MRC UT Austin, TX 77005 USA. He is now with Department of Physics, University of Maryland, College Park, MD 20742 USA, (samaresh@physics.utexas.edu)

Equivalently, given the opposite relation between the direction of electron flow and charge current flow and between electron spin orientation and associated magnetic moment, a current flow in the positive $x$ direction induces a magnetization oriented in the negative $y$ direction on the surface of the TI. The relative spin polarization of the TI surface electrons and a polarized ferromagnetic metallic contact can affect the current-voltage relationship for current flow through the contact, resulting in a magnetoresistance [4].

Low temperature magneto-transport measurements have been performed by several groups using stack(s) of molecular beam epitaxy (MBE) grown ferromagnetic metal (FM) layers either on TI flakes or on MBE-grown TIs [11-14]. These groups have used $Bi_2Se_3$ or BiSbTe TIs and Fe, Co or Py FMs. Recently, Dankert *et al.* probed the spin polarization on the $Bi_2Se_3$ surface using Co FM electrodes at room temperature (RT) [15]. The latter group showed a change in voltage ($\Delta V$) on the scale of a few μV between magnetization orientations of a Co FM layer for a surface current density of ~100 μA/μm [15].

In this work, we have overlayed an Fe FM on exfoliated $Bi_2Te_3$ TI layers with and without an evaporated-oxide layer in between, and on exfoliated $Bi_2Se_3$ without an evaporated oxide layer, although a residual native oxide also appears likely for the considered devices absent the evaporated oxide layer. We have demonstrated large magnetoresistances at temperatures ranging from 100 K to RT, with voltage changes of a few mV with lower current densities and associated greater magnetoresistance than previously reported (up to a few percent in the latter two of these systems).

## II. Experiments

The devices used for our magneto-transport measurements incorporate vertical stacks of FM Fe on TI and of nonmagnetic Au on TI, at various lateral inter-stack spacings, as illustrated in Fig. 1. A thin TI film was exfoliated mechanically from a bulk single crystal of $Bi_2Te_3$ or of $Bi_2Se_3$ using the standard "Scotch tape" method on a $SiO_2$/Si substrate. The $SiO_2$ layer is 300 nm thick. The FMs were deposited by e-beam evaporation and patterned by e-beam lithography. For some $Bi_2Te_3$ devices, ~ 2 nm thick thin oxide layer ($SiO_2$) was evaporated before Fe layer was deposited. The Au contacts to the TI and the Au caps on Fe electrodes were fabricated by a second lithographic and lift-off process. Actual Au and Fe contacts on an exfoliated flake of $Bi_2Te_3$ are shown in Fig. 1(b). Raman spectroscopy showed the expected in-plane vibrational mode ($E_g^2$), and two out-of-plane modes ($A_{1g}^1$ and $A_{1g}^2$) for $Bi_2Te_3$ and $Bi_2Se_3$, respectively [16, 17]. Atomic force microscopy (AFM) images were obtained to estimate the thickness of the TI and Fe layers. The thicknesses of the TI flakes ($Bi_2Te_3$ and $Bi_2Se_3$) range from 180 nm to 100 nm, such that there should be no distortion of the surface state band structure by inter-surface coupling. The Fe layer thickness is ~50 nm.

Two terminal current-voltage (*I-V*) magneto-transport measurements were performed, as schematically illustrated in Fig. 1(a), using a Quantum Design physical property measurement system at temperatures between 100 K and RT. The magnetization easy-axis of the Fe bar is in-plane, along its long axis normal to the direction of current flow. The definition of positive and negative current flow, and positive/up (↑) and negative/down (↓) magnetization *M* of the Fe bar and the TI surface are consistent with the coordinate axes shown in Fig. 1. The magnetic orientation of the bar was set using an external in-plane magnetic field *prior* to each individual current-voltage (*I-V*) trace. Two sets of *I-V* data were collected, one for magnetization up (↑) (net electron spin down (↓)) in the Fe bar with respect to the *y* axis and another for magnetization down (↓) (net electron spin up (↑)). To isolate the magnetoresistive effects, the voltage difference $\Delta V$ is obtained by subtracting these two sets of *I-V*, $\Delta V = V_{H\uparrow} - V_{H\downarrow}$, as illustrated in Fig. 1(c).

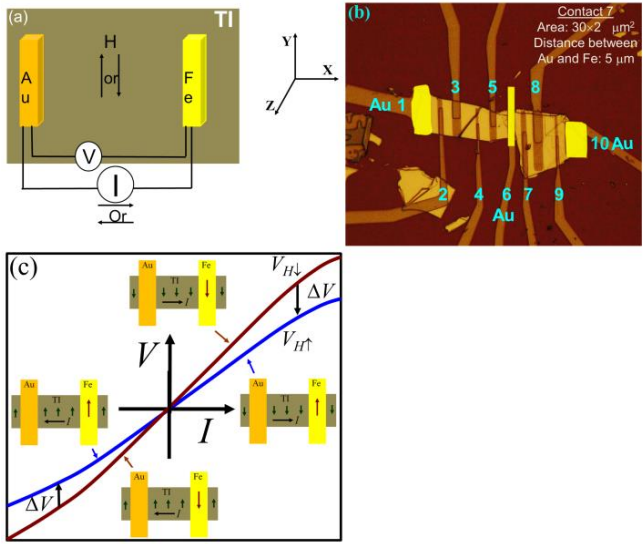

Fig. 1. (a) Schematic of current and voltage probe configuration between Au and Fe contacts placed in parallel on a TI flake. The magnetization of the Fe bar is aligned along the length of the bar using an external magnetic field before each current-voltage (*I-V*) sweep. Fe in-plane magnetization and the direction of the applied current are orthogonal to each other. (b) An experimental device with Au contacts (1, 6 and 10; highlighted in yellow on the image) and Fe (2, 3, 4, 5, 7, 8 and 9) contacts fabricated on an exfoliated $Bi_2Te_3$ flake [18]. The Fe layer is of 50 nm thick, as estimated from AFM measurements. Au Contacts 6 and Fe Contact 7 were used for the magneto-transport measurements for this device. The width of the Fe bar along the transport direction (*x*) is about 2 μm. The channel length between the Au and Fe contacts is approximately 5 μm. The effective channel width normal to the transport direction for this irregularly shaped TI flake is about 15 μm. (c) Schematic illustration of extraction of $\Delta V$ from *I-V* plots (with exaggerated magnetoresistive effects here for clarity) for up (↑) and down (↓) Fe *magnetization* alignments. (The $\Delta I$ of the *Results and Discussion*, would correspond to half the horizontal shifts between the curves in this figure and be of opposite sign.) The orientation of the net *magnetic moment* of electrons on the TI surface for the positive and negative current sweep is also shown with smaller arrows on the TI. The voltage drop/current flow is measured from the Au to the Fe contact.

### III. RESULTS AND DISCUSSIONS

Among our samples, current-voltage (*I-V*) characteristics varied from ohmic to a more tunneling-like non-ohmic behavior with higher resistance at low currents/voltages even for some devices without an evaporated oxide. Only the devices with observable non-ohmic characteristics exhibited magnetoresistance. In addition, when we exposed $Bi_2Te_3$/Fe and $Bi_2Se_3$/Fe devices without evaporated oxide to currents exceeding approximately 100 μA, the contacts exhibited a permanent shift to low-resistance ohmic behavior and the magnetoresistance was lost. Although these latter devices did not have an evaporated oxide, formation of a thin native oxide at the Fe/TI interface and the subsequent breakdown thereof at high currents, at least locally, may be responsible for these observed behaviors. Similarly, devices with initially ohmic *I-V* characteristics may not have had a significant native oxide or perhaps an incomplete one.

This association of magnetoresistance with non-ohmic contact resistance is suggestive of a ferromagnet/oxide/TI contact analog of tunneling magneto resistance TMR but with some qualitative differences. In Fe, the spin up density of states $N_{Fe,\uparrow}$ and the spin down density of states $N_{Fe,\downarrow}$ in the vicinity of the Fermi level are different due to ferromagnetism. In contrast, however, on the TI surface with spin-momentum locked electrons, the density of states remains fixed, but current flow will create changes in the spin up electron density $\Delta n_{TI,\uparrow}$ and spin down electron density $\Delta n_{TI,\downarrow}$ that are necessarily of equal magnitude and opposite sign, $\Delta n_{TI,\uparrow} = -\Delta n_{TI,\downarrow}$, for a fixed total electron density. Consistent with the original TMR model of Julliere [19], we assume that the tunnel transmission through the oxide is randomized with respect to momentum while spin is conserved. With the above three conditions, there will be a charge current *change* $\Delta I$ compared to what otherwise would be expected (again taken as positive flowing into the magnet) associated with electron injection from the TI to the Fe or, as the case may be, reduction in electron extraction to the TI from the Fe in proportion to the changes in the electron densities, $\Delta n_{TI,\uparrow}$ and $\Delta n_{TI,\downarrow}$, and the respective spin densities of states in the Fe, $N_{Fe,\uparrow}$ and $N_{Fe,\downarrow}$: $\Delta I \propto -q(\Delta n_{TI,\uparrow} N_{Fe,\uparrow} + \Delta n_{TI,\downarrow} N_{Fe,\downarrow}) = q\Delta n_{TI,\uparrow}(N_{Fe,\downarrow} - N_{Fe,\uparrow})$. For an up magnetization of the Fe bar and an associated overall majority spin down electron concentration, the associated spin densities in the vicinity of the Fermi level at least effectively are such that $N_{Fe,\downarrow} > N_{Fe,\uparrow}$ [20-22]. If the overall current flow $I$ is positive such that electrons are moving in the negative $y$ direction, $\Delta n_{TI,\uparrow}$ will be positive for the spin-momentum locked surface states, and $\Delta I$ will also be positive. Therefore, the overall current magnitude will increase and the resistance will decrease. If the sign/direction of overall current flow is flipped, so will be the sign of $\Delta I$, and again the overall current magnitude will increase and the resistance will decrease. However, if the magnetization of the Fe bar is flipped, $\Delta I$ flow will oppose the overall current flow $I$, and the resistance will increase. This expectation is depicted in Fig. 1(c) in terms of the orientations of the magnetizations of the Fe bar and TI surface (both opposite their respective majority spin concentrations), and results in a negative $\Delta V = V_{H\uparrow} - V_{H\downarrow}$ for positive current and a positive $\Delta V$ for negative current. Finally, given the oxides employed, we have not considered additional spin-filtering by the tunnel oxide itself.

For the $Bi_2Te_3$/Fe device of Figure 1(b), which was fabricated without an $SiO_2$ interlayer, we observed magnetoresistive effects on $\Delta V$ from 100 K to RT, as shown in Fig. 2. The sign of $\Delta V$ and the change thereof with current direction here, as well as for all examples to follow, is consistent with the above expectations. The $\Delta V$ are on the scale of 100 μV at RT to 500 μV at lower temperatures at ±60 μA (~4 μA/μm). This $\Delta V$ is substantially greater than previously reported even while at lower current densities [15]. The corresponding magnetoresistance, taken as $|\Delta V/(V_{H\uparrow} + V_{H\downarrow})|$ for fixed current, is about 0.1 % at room temprature and in the 0.3 % range at lower temperatures at ±60 μA. (See., e.g., the inset of Fig 3 for the room temperaure $I$-$V$ characeristic required for the denominator.)

Fe and $Bi_2Te_3$ layers, and for $Bi_2Se_3$, again with Fe and Au contacts but without an evaporated oxide, and found substantially larger magnetoresistances. Fig. 3 shows $\Delta V$ vs. applied current at RT for three devices: the $Bi_2Te_3$-based device without evaporated oxide of Figs. 1(b) and 2 for reference, a $Bi_2Te_3$-based device with an evaporated interfacial $SiO_2$ layer, and a $Bi_2Se_3$-based device without evaporated oxide. The inset shows the corresponding $I$-$V$ characteristics for these three devices. The advantage of introducing the evaporated oxide for the $Bi_2Te_3$-based device is evident, with the magnetoresistance now increasing to ~4%. However, the effect on magnetoresistance for the Fe/$Bi_2Se_3$-based device without evaporated oxide (with larger $\Delta V$ but also larger $V$ for a given current) is not much lower, perhaps again due to the presence of a native oxide. Moreover, $Bi_2Se_3$ may be a model TI for RT application [23], and one in which the deposition of Fe directly on $Bi_2Se_3$ does not affect the topological surface states [24]. In contrast, Fe deposition directly on $Bi_2Te_3$ may form an Fe-rich FeTe compound at the interface, degrading the quality of interface states [25] and, thus, magnetoresistance, which further attests to the advantage of any interlayer oxide, native or evaporated, that may be present for the $Bi_2Te_3$-based systems.

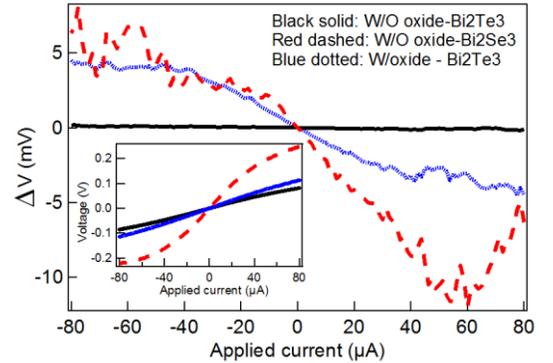

Fig. 3. Plots of $\Delta V$-$I$ for three different devices at room temperature (positive to negative current sweeps): the $Bi_2Te_3$,-based device without an evaporated oxide (solid curve, black online) of Fig. 1(b) and 3; a $Bi_2Te_3$-based device with an evaporated $SiO_2$ oxide (dotted curve, blue online) and with an effective intercontact channel dimension of approximately 20 μm long by 15 μm wide with a 2 μm Fe contact within the transport direction; and a $Bi_2Se_3$-based device without an evaporated oxide (dashed curve, red online) and with an effective intercontact channel dimension of approximately 25 μm long by 5 μm wide with a 2 μm Fe contact within the transport direction. The inset shows the corresponding $I$-$V$. An in-plane magnetic field of ± 5T was applied for the pre-alignment of the Fe bar in each case.

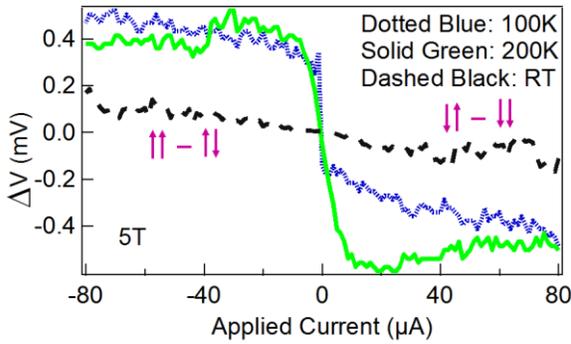

Fig. 2. $\Delta V$-$I$ (positive to negative current sweeps) for a temperature range between 100K and RT (dotted and blue online: 100K; solid and green online: 200K; and dashed and black online: RT). An in-plane magnetic field of ± 5T was applied for the pre-alignment of the Fe bar. Each pair of the arrows shows the relative alignment of the magnetization of the TI surface and the Fe bar.

We then performed similar measurements for $Bi_2Te_3$-based devices but with an evaporated oxide layer between the

The observed magnetoresistive effects in these samples are highly reproducible and independent of measurement order. Fig. 4 displays $I$-$\Delta V$ for the evaporated-oxide $Bi_2Te_3$ device, with multiple runs including both positive-to-negative and negative-to-positive current sweeps, and pre-aligning the magnet and then resetting it to zero and pre-aligning again. As a control we also measured current flow between Au contacts and were unable to discern any magnetoresistance.

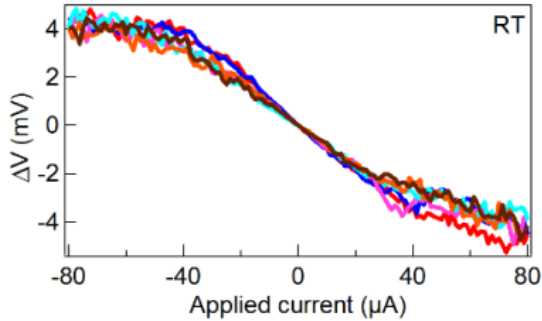

Fig. 4. $\Delta V$ as a function of applied positive to negative current sweep for an Fe/SiO$_2$/Bi$_2$Te$_3$ device at RT for multiple runs including two consecutive runs with pre-aligning the Fe bar initially and then resetting the magnet to zero and again pre-aligning it, and then repeated for applied negative to positive current sweeps. All runs show similar behavior, exhibiting the reproducibility of these measurements.

## IV. CONCLUSION

Magnetoresistance has been observed for nominally Fe/Bi$_2$Te$_3$, Fe/Bi$_2$Se$_3$ and Fe/evaporated-SiO$_2$/Bi$_2$Te$_3$ contacts from 100 K up to RT, although it appears that a native oxide is present even in the absence of the evaporated oxide. This magnetoresistance is consistent with qualitatively expected dependencies on the relative orientations of the magnetization of the Fe bar and the spin of the spin-momentum locked electron on the TI surface. The observed magnetoresistance is substantially larger than previously reported, and is particularly enhanced when an evaporated SiO$_2$ layer is introduced between Fe and Bi$_2$Te$_3$, and when Bi$_2$Se$_3$ is used instead of Bi$_2$Te$_3$.